\documentclass{article}
\usepackage{spconf,amsmath,graphicx}
\usepackage{makecell}
\usepackage{float}
\usepackage{stfloats}
\usepackage{booktabs}

\usepackage[colorlinks=true,citecolor=green]{hyperref}
\usepackage{cleveref}
\usepackage{caption}
\usepackage{subcaption}
\usepackage{adjustbox}
\usepackage{subfiles}
\usepackage{amssymb}
\usepackage{pifont}
\usepackage{multirow}
\usepackage{xcolor}

\newcommand{\cmark}{\ding{51}}%
\newcommand{\xmark}{\ding{55}}%

\makeatletter
\renewcommand\section{\@startsection {section}{1}{\z@}%
                                  {-2.5ex \@plus -1ex \@minus -.2ex}%
                                  {1.75ex \@plus.2ex}%
                      {\normalfont\Large\bfseries}}
\renewcommand\subsection{\@startsection{subsection}{2}{\z@}%
                                     {-1.5ex\@plus -1ex \@minus -.2ex}%
                                     {1.ex \@plus .2ex}%
                              {\normalfont\large\bfseries}}
\renewcommand\subsubsection{\@startsection{subsubsection}{3}{\z@}%
                                     {-3.25ex\@plus -1ex \@minus -.2ex}%
                                     {1.5ex \@plus .2ex}%
                          {\normalfont\normalsize\bfseries}}
\renewcommand\paragraph{\@startsection{paragraph}{4}{\z@}%
                                    {3.25ex \@plus1ex \@minus.2ex}%
                                    {-1em}%
                        {\normalfont\normalsize\bfseries}}
\renewcommand\subparagraph{\@startsection{subparagraph}{5}{\parindent}%
                                    {3.25ex \@plus1ex \@minus .2ex}%
                                    {-1em}%
                          {\normalfont\normalsize\bfseries}}
\makeatother

\title{Optimizing alignment of speech and language latent spaces for end-to-end speech recognition and understanding}
\name{\begin{tabular}{c}Wei Wang$^{1,2,*}$ \thanks{$^*$Work done during an internship at Microsoft.},  Shuo Ren$^2$, Yao Qian$^2$, Shujie Liu$^2$, Yu Shi$^2$, Yanmin Qian$^1$, Michael Zeng$^2$ \end{tabular} }
\address{
$^1$ MoE Key Lab of Artificial Intelligence, AI Institute\\
X-LANCE Lab, Department of Computer Science and Engineering, Shanghai Jiao Tong University\\
$^2$ Microsoft Corporation}

\begin{document}
\ninept

\maketitle



\begin{abstract}
The advances in attention-based encoder-decoder (AED) networks have brought great progress to end-to-end (E2E) automatic speech recognition (ASR). 
One way to further improve the performance of AED-based E2E ASR is to introduce an extra text encoder for leveraging extensive text data and thus capture more context-aware linguistic information. However, this approach brings a mismatch problem between the speech encoder and the text encoder due to the different units used for modeling.
In this paper, we propose an embedding aligner and modality switch training to better align the speech and text latent spaces. 
The embedding aligner is a shared linear projection between text encoder and speech encoder trained by masked language modeling (MLM) loss and connectionist temporal classification (CTC), respectively. The modality switch training randomly swaps speech and text embeddings based on the forced alignment result to learn a joint representation space.  
Experimental results show that our proposed approach achieves a relative 14\% to 19\% word error rate (WER) reduction on \textsc{Librispeech} ASR task. We further verify its effectiveness on spoken language understanding (SLU), i.e., an absolute 2.5\% to 2.8\% F1 score improvement on \textsc{Snips} slot filling task.
\end{abstract}
%


\section{Introduction}
\label{sec:intro}
Since the emergence of end-to-end~(E2E) models, the automatic speech recognition~(ASR) pipeline has been greatly simplified, and ASR tasks can be accomplished with a unified model architecture~\cite{chiu2018stateoftheart, Zeyer2018}. The most commonly adopted E2E ASR architectures are attention-based encoder-decoder models~\cite{chorowski2015attentionbased, kim2017joint}. In these models, the encoder plays an essential role, which converts the acoustic information to context-aware linguistic features. The decoder then generates the formal text output based on the linguistic features.  

In recent years, many attempts have been made to assist the encoder in learning context-aware linguistic information with large text corpora~\cite{tang2021general, renduchintala2018multimodal,9003873, huang2020leveraging,wang2020bridging}. 
An E2E ASR model with an extra text encoder network is a commonly used architecture to integrate more linguistic information into the ASR encoder. \cite{9003873} incorporates a smoothed L1 loss with a multi-stage training scheme and trains the text encoder and speech encoder to match their output to each other, pushing the speech embeddings closer to the text embedding space.

Instead of casting explicit constraints, in \cite{tang2021general}, the text encoder shares part of its layers with the speech encoder and is trained on large text corpora with an extra text denoising autoencoder task. Under this multi-task framework, the shared encoder is capable of encoding both speech and text embedding from their corresponding encoder, mitigating the mismatch between the embedding from the shared encoder and text decoder. \cite{renduchintala2018multimodal} examins different synthetic input generation schemes for text data and concludes that repeating phoneme input by their relative duration to each other achieves the best performance. 
Rather than introducing another encoder for text data, \cite{chen2018modular} utilizes traditional acoustic and language modeling techniques in hybrid systems. An acoustic-to-phoneme module and a phoneme-to-word module are trained separately but decoded in an E2E fashion.
\cite{chen2021injecting, jia2019leveraging, Laptev_2020} adopt text-to-speech~(TTS) technique to utilize the extensive text corpora to generate labeled speech and improve the generalization ability of ASR models without a modification to the model structure. 

Following previous work, we also leverage an extra text encoder network to learn contextual representations from a large text corpus in this paper. While the framework is similar to the multi-modality framework used in \cite{tang2021general}, we propose an embedding aligner to reduce the mismatch between speech and text embeddings,  which was not considered in these earlier studies. Under this refined framework, the text encoder trained with MLM target becomes a simplified phoneme bert~\cite{sundararaman2021phoneme}, injecting phoneme information into the embedding aligner. The speech encoder trained with phoneme CTC target learns from the aligner and encodes speech into embeddings that are easier to align with those from the text decoder. Meanwhile, dot product is replaced with Euclidean distance to calculate the pairwise distance between input embedding and embedding aligner for making the embeddings from speech and text further closer. \cite{renduchintala2018multimodal} only took advantage of the extensive text data and the text encoder to mimic speech input for data augmentation. \cite{9003873} performed experiments under a low-resource setup and adopted a different target to enforce the speech-language alignment. Inspired by the code-switch methods in machine translation \cite{yang2020codeswitching, lin2021pretraining} for better alignment, we extend it to a modality switch training~(MST) method to swap speech and text embedding for enhancing speech-text alignment in ASR training, based on the forced alignment result. We conduct experiments on \textsc{Librispeech} ASR tasks and \textsc{SNIPs} slot filling (SF) task of spoken language understanding (SLU). The experimental results from both tasks show our proposed methods are promising. 

Our contributions can be summarized as: 1) We propose an embedding aligner method to explicitly align the speech and text hidden spaces by sharing the same weights optimized by both the MLM and CTC losses; 2) Euclidean distance is employed to make the speech and language spaces tied closer; 3) A modality switch training (MST) method is proposed to fuse the embeddings generated from speech and text encoders to enhance the alignment between them; 4) The effectiveness of our proposed methods has been demonstrated on both ASR and SLU tasks, i.e., significantly reduce the WER of ASR on \textsc{Librispeech} and improve the F1 score of SF on \textsc{SNIPs}.

\begin{figure*}[tp]
  \centering
  \includegraphics[width=\linewidth]{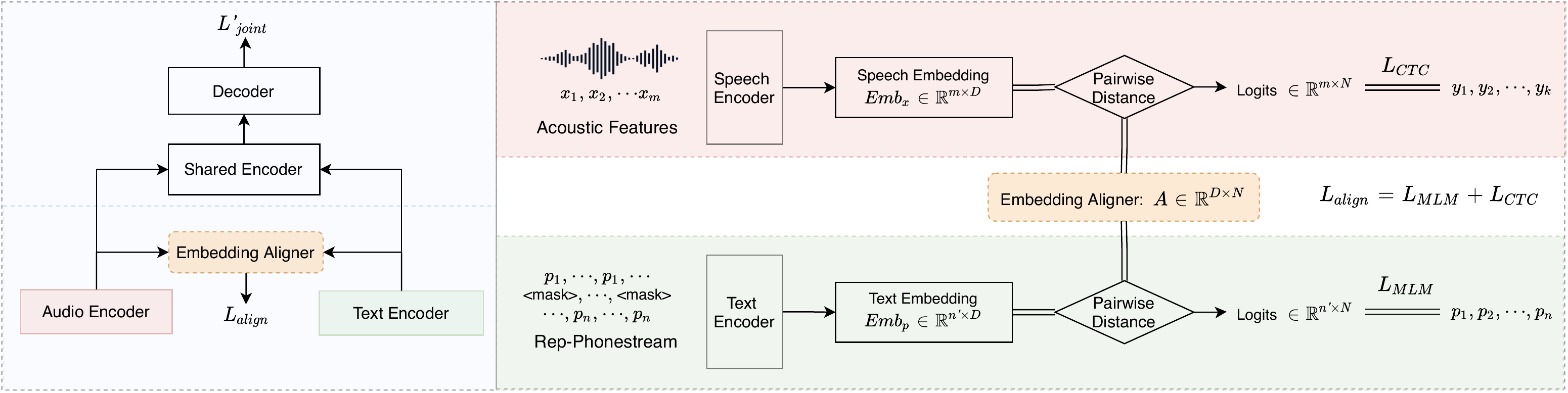}
  \caption{The proposed framework to enhance the speech-text embedding alignment. The left part shows the overall structure of the framework. We adopt joint CTC-attention multi-task loss in the decoder side and attach our proposed alignment loss on the encoder side. The right part shows the details of the encoders and the embedding aligner. We calculate the pairwise distance between embeddings from two encoders and the embedding aligner. The output logits are used for MLM and CTC target computation. Two pairwise distance metrics are compared: (1) Dot product (2) Euclidean distance. }
  \label{fig:aligner}
\end{figure*}



\section{Methodology}
\label{sec:methodology}
The whole network structure is illustrated in the left part of Fig. \ref{fig:aligner}. For the speech input, an audio encoder is used to extract speech embeddings, and the shared encoder is used to learn context-aware linguistic features, followed by a decoder to generate the ASR output. For the text data, the masked text input is processed with the text encoder, the shared encoder and the shared decoder to recover the original text data. 
To deal with the mismatch problem between the audio encoder and the text encoder, in this section, we introduce two approaches, the embedding aligner, and the modality switch training, to explicitly align the hidden spaces of speech and text.

\subsection{Embedding Aligner}
\label{subsec:embedding_aligner}
As shown in the right part of Fig.\ref{fig:aligner}, we trained the text encoder with MLM target and the speech encoder with CTC targe on a text corpus. To extract linguistic information from phoneme sequences and to model acoustic information with phoneme tokens, both MLM and CTC targets adopt phoneme representations of text sequences. 

The embedding aligner is a linear projection shared between MLM and CTC targets, which is used for calculating the logits in the softmax layer. Denote the dimension of speech and text embedding as $D$ and the size of the phoneme dictionary as $N$. The embedding aligner is denoted as $A$~($A\in\mathbb{R}^{D \times N}$), representing the learnable embedding for each entrance in the phoneme dictionary. 
To deal with the length mismatch between speech and text, we model phoneme relative duration by Rep-Phonestream~\cite{renduchintala2018multimodal}, and repeat phonemes by their relative duration to each other. 
Denote $P = (p_1, p_2, \cdots, p_n)$ as the phoneme representation of a text sequence with $n$ phoneme tokens, where 20\% of the phonemes are randomly replaced with $<$mask$>$ symbols. After repeating phonemes, we have $P_{rep} = (p_1,\cdots, p_1, p_2, \cdots, p_2, p_n, \cdots, p_n)$, whose length is denoted as $n'$. The phoneme MLM target can be calculated as:
\begin{align}
    Emb_p &= \operatorname{Text-Encoder}(P_{rep}) \\
    \label{eq:softmax_mlm} \hat{P} &= \operatorname{Softmax} (E_p \cdot A) \\
    L_{MLM} &= \operatorname{MLMLoss} (\hat{P}, P)
\end{align}
where $Emb_p$ is the embedding of $P$ extracted by text encoder. Denote $X = (x_1, x_2, \cdots, x_m)$ as the acoustic features of a speech sample with $m$ frames. Its corresponding phoneme transcripts is denoted as $Y = (y_1, y_2, \cdots, y_k)$. The phoneme CTC target can be calculated as:
\begin{align}
    Emb_x &= \operatorname{Speech-Encoder}(X) \\
    \label{eq:softmax_ctc} \hat{Y} &= \operatorname{Softmax} (E_x \cdot A) \\
    L_{CTC} &= \operatorname{CTCLoss} (\hat{Y}, Y)
\end{align}
where $Emb_x$ is the embedding of $X$ extracted by speech encoder. Note that in Eq.\eqref{eq:softmax_mlm} and Eq.\eqref{eq:softmax_ctc}, the logits in the softmax layer is calculated as the pairwise dot product distance between the input embedding and each entrance of the embedding aligner. We replace the dot product distance with euclidean distance to cast a stronger restriction on the pairwise distance between input embedding and embedding aligner in both direction and scale. Concretely,  Eq.\eqref{eq:softmax_mlm} and Eq.\eqref{eq:softmax_ctc} are rewritten as:
\begin{align}
    \hat{P} &= \operatorname{Softmax} (\operatorname{Pairwise-Euclidean} (E_p, A)) \\
    \hat{Y} &= \operatorname{Softmax} (\operatorname{Pairwise-Euclidean} (E_x, A))
\end{align}
It should be noted that, $\hat{P}$ and $\hat{Y}$ share the same weight $A$, and the two objectives $ L_{MLM}$ and $L_{CTC}$ try to push both the hidden states $E_p$ and $E_x$ to $A$.
The loss of embedding aligner becomes:
\begin{equation}
    L_{align} = L_{MLM} + L_{CTC}
\end{equation}

Both dot product and Euclidean distance can measure the similarity between two vectors. These two similarity measurements are equivalent if the vectors are unit-length, but the dot product is proportional to the vector length. We conducted experiments with dot product and length normalization for the embeddings $E_p$ and $E_x$, but cannot get any promising improvement.

For the decoder part, we adopt a joint CTC attention training framework in \cite{kim2017joint} and subwords as modeling units for text~\cite{kudo2018subword} denoted as $L'_{joint}$. Note that this multitask target is applied for both paired speech data and unpaired text data. Combining encoder and decoder, the loss under our proposed framework is:
\begin{equation}
    L = \alpha L_{align} + (1 - \alpha) L'_{joint}
\end{equation}
where $\alpha$ is the weight of the embedding alignment loss.

\subsection{Modality Switch Training}
\label{subsec:modality_switch_training}
Aligning the embedding space of input and output sequence by randomly replacing input tokens with their corresponding output tokens has been proved effective in multi-lingual machine translation ~\cite{ yang2020codeswitching,lin2021pretraining}. 
The success of this approach in machine translation can be ascribed to two reasons: (1) The input and output spaces are discrete and share the same token list, which reduces the mismatch between the embeddings of source and target languages. (2) Most input and output tokens correspond one by one. Thus the replacement won't make a big difference in sequence length. Regarding the ASR tasks, for the first one, the mismatch between input and output spaces can be minimized by our proposed embedding aligner. For the second, given a pair of speech and phoneme sequences, we repeat each phoneme by its duration according to the force alignment results, and generate a new phoneme sequence whose length is exactly the same as speech frames. We compare two strategies regarding the swapping methods: phoneme-unaware and phoneme-aware, as illustrated in Fig.\ref{fig:MST}:

\begin{figure}[H]
  \centering
  \includegraphics[width=\linewidth]{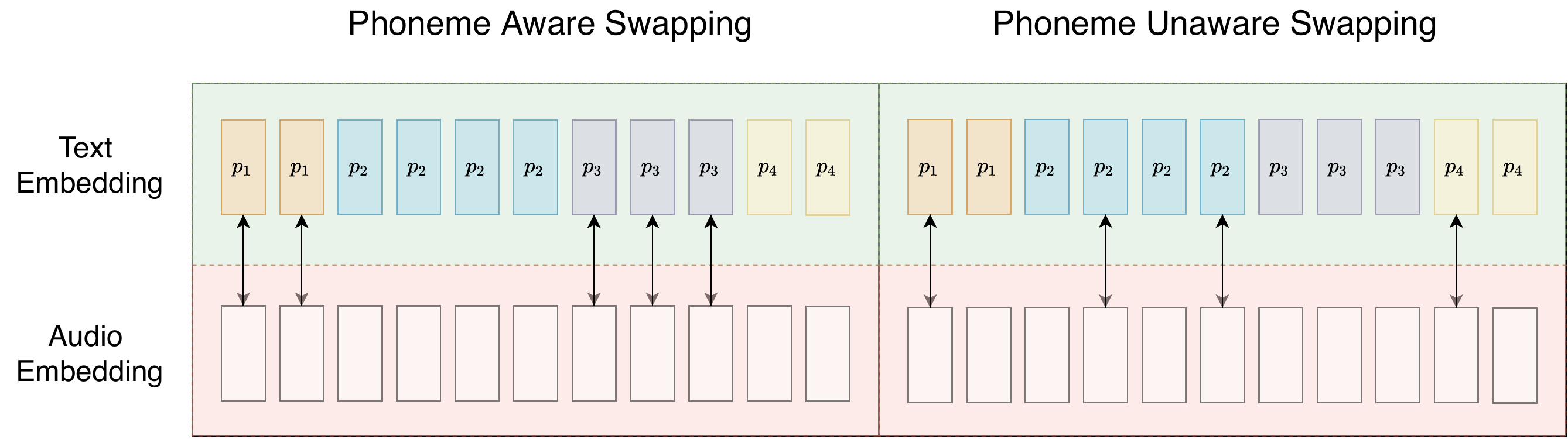}
  \caption{Modality switch training with phoneme aware and phoneme unaware strategy: (1) phoneme-aware: embedding swapping is performed on spans of frames that correspond to the same phoneme token in forced alignment. (2) phoneme-unaware: random frames are selected to perform the embedding swapping regardless of their correspondence to phonemes. MST is only applied when training the model on speech data with transcription, where the transcripts are used as inputs for the text encoder. 10\% of phonemes or frames are randomly selected for MST. }
  \label{fig:MST}
\end{figure}


\section{Experiments}
\label{sec:experiments}

\subsection{Experiment Setup}
\label{sec:exp_setup}
\subsubsection{ASR task}
ASR experiments are carried out on 960 hours of \textsc{Librispeech} dataset and a text corpus, which contains 14500 public domain books and comes with \textsc{Librispeech} for language model training.
We adopt 10 layers of speech encoder, 4 layers of text encoder, 2 layers of shared encoder and 6 layers of the decoder with 2048 hidden units. Each layer is a Transformer block with 8 heads of 64 dimension self-attention layer~\cite{vaswani2017attention}. For multitask learning~(MTL), the weight for CTC and attention is set to 0.3 and 0.7. We use an 80-dimensional log Mel-filterbank with 25ms window length computed every 10ms as inputs of speech encoder. Spec-augment with policy LD~\cite{Park_2019} is applied during training. The model is trained on both unpaired text data and pair speech data alternately for each batch. The Adam~\cite{kingma2014adam} optimizer is adopted with 0.001 initial learning rate and 20,000 warmup steps. All models are trained until convergence. 

For the text encoder, the input is position-dependent phonemes generated with Rep-Phonestream strategy for the text corpus. The force alignment results obtained from the TDNN chain model in Kaldi~\cite{kaldi,peddinti2015time} are used as inputs to text encoder for transcripts of \textsc{Librispeech} audios. And the relative duration of phoneme estimated from the force alignment results for speech data are used to repeat phonemes in unpaired text data. We use the official \textsc{Librispeech} lexicon to perform grapheme-to-phoneme transduction (G2P). There are over 300 position-dependent phonemes in the phoneme dictionary. The number of modeling units (BPE) for text sequence in the decoder is 10,000~\cite{kudo2018subword}. Experiments are carried out with ESPnets toolkit~\cite{watanabe2018espnet}.

\vspace{-.5em}

\subsubsection{SLU task}

 SLU experiments are carried out on \textsc{Snips} dataset and perform a slot filling~(SF) task. We employ the same data split as in Audio SNIPS corpus, which contains synthesized multi-speaker utterances for SNIPS and was used for SF task in SUPERB Benchmark~\cite{yang2021superb}. The SF task requires a model to extract slot type and slot value pairs directly from speech inputs. The SF task here is reformulated as a sequence-to-sequence problem by predicting a sequence in which the slot value is surrounded by slot type boundaries. For example, a slot value and slot type pair ``served\_dish : maple syrup'' is converted to ``B-served\_dish maple syrup E-served\_dish''. We also follow SUPERB to use 
 Character Error Rate (CER) for slot value and F1 score for slot type to evaluate the performance of SF task.
 We adopt Transformer with 12 layers of encoder and 2 layers of decoders for SLU task. The CTC weight in MTL is set to zero in this task.

\subsection{Experiment Results and analysis}
\label{sec:exp_results}
\begin{table*}
\centering
\renewcommand{\arraystretch}{1.3}
\begin{tabular}{c|c|c|c|cccc}
\hline \hline\noalign{\smallskip}
\multirow{2}{*}{Multi-Modal} & Num of Shared Layers & Embedding Aligner  &  \multirow{2}{*}{Modality Switch Training}  & \multicolumn{2}{c}{Dev} & \multicolumn{2}{c}{Test} \\
& in Encoder & (Distance Metrics) & & clean & other & clean & other  \\
\noalign{\smallskip}\hline
\xmark~\cite{tang2021general}  & N/A &  N/A  &  N/A      & 3.5  &  8.1 & 3.7  &  8.1 \\
\xmark  &  N/A  & N/A &  N/A  & 3.3  &  8.0 & 3.6  &  8.0 \\
\noalign{\smallskip}\hline
\cmark~\cite{tang2021general} & 6 &  N/A  & N/A  &  3.0  & 7.4  &  3.3 & 7.6  \\
\cmark~\cite{tang2021general} & 0 &  N/A  & N/A  &  3.0  & 7.4  &  - & -  \\
\cmark  & 2 & N/A  &  N/A   & 3.1  & 7.6  & 3.5  & 7.3  \\
\noalign{\smallskip}\hline
\cmark  &  2 & Dot Product   &   N/A & 2.9 & 7.2 & 3.1 & 7.2   \\
\cmark   & 2 & Euclidean   &   N/A  & 2.7 & 7.1 & 3.0 & 7.0  \\
\cmark  & 4  &  Euclidean    &   N/A  & 2.7 & 7.2 & 2.9  & 7.0 \\
\cmark   & 2 & Euclidean   &   Phoneme Unaware &  2.8 & 7.1 & 3.1 & 6.9   \\
\cmark   & 2 &  Euclidean    &  Phoneme Aware & \textbf{2.7} & \textbf{6.9} & \textbf{2.9} & \textbf{6.8}  \\
\noalign{\smallskip}\hline\hline
\end{tabular}
\caption{Performance comparison (WER\%) of different setups}
\label{tab:performance}
\end{table*}

The ASR performance for different system setups is shown in Table \ref{tab:performance}. It also lists the performance of reference systems from ~\cite{tang2021general}, which leveraged text data in a multi-modality transformer framework. We reproduced the results in ~\cite{tang2021general} and used them as our baseline systems. Note that the joint-CTC-attention framework was not adopted in \cite{tang2021general} since the text outputs modeled with subword units were not necessarily shorter than the text inputs modeled with phoneme units. We repeated the phoneme inputs to similar lengths as speech inputs, enabling the application of an extra CTC loss on the decoder side. The corresponding WERs are shown in the second row and fifth row. We got the same findings as those reported in ~\cite{tang2021general}, i.e., adding a text encoder trained with the phoneme inputs is beneficial to ASR performance and the number of shared layers between audio encoder and text encoder plays a marginal impact on WER.

\vspace{-.5em}

\subsubsection{Embedding aligner}
The effectiveness of our proposed embedding aligner is shown in the bottom part of Table \ref{tab:performance}. Embedding aligner can improve the ASR performance by relative 1.3\% to 11.4\% WER reductions on different evaluation sets. In addition, replacing the dot product with Euclidean distance as the similarity measurement among the embeddings from both encoders and the embedding aligner can further improve the performance by a relative 3.6\% WER reduction, averaged over four evaluation sets. We observe that the relative improvement on clean sets is larger than on other sets. 
A possible explanation is that the matched condition between training and testing on the clean set makes the speech encoder relatively easy to generate embeddings closer to its correct entrance of the embedding aligner, and vice versa.

\vspace{-.5em}

\subsubsection{Modality switch training}
We compared the phoneme-aware and phoneme-unaware strategies of MST. The results are shown in the last two rows of Table~\ref{tab:performance}. Phoneme-unaware MST did not improve the performance. One possible reason is that among a span of frames that corresponds to the same phoneme, only a few frames were replaced by the text embeddings. Thus the encoder was still able to utilize speech embedding from other frames to perform the phoneme recognition. Under such circumstances, phone-unaware MST has similar effects as the time-warp strategy in spec-augment, which has already been employed in our system. With phoneme-aware MST, the encoder must learn to use text embeddings of the entire span of frames that corresponds to the selected phoneme to reconstruct the phoneme. The phoneme-aware MST is similar to the semantic mask proposed in \cite{wang2020semantic} except that the masked embeddings are substituted with corresponding text embeddings. Therefore, the phoneme-aware MST can further improve the robustness of the embedding alignment and obtain a slight improvement on both dev-other and test-other sets.

\vspace{-1em}
\subsubsection{Text data usage}
Since in our system, text data for training \textsc{Librispeech} LM was used as multi-modality input, for a fair comparison, we further conducted experiments that trained an LM on the same text data and applied shallow fusion with a single-modality ASR model. The results are shown in Table \ref{tab:lm-compare-result}. It indicates that utilizing the text data to train an embedding aligner that optimizes the speech-text alignment on the encoder side yields better results than a shallow fusion with LM trained on the same text data on the decoder side. It also shows that our proposed method of using text data is complementary to LM shallow fusion, and the performance of the multi-modality trained model can be further improved by LM rescoring. 

\begin{table}[htbp]
  \centering
  \begin{tabular}{l  | c c | c c}
    \toprule
    \multirow{2}{*}{System} & \multicolumn{2}{c|}{Dev} & \multicolumn{2}{c}{Test} \\
     & clean & other & clean & other \\
    \midrule
    Single-Modal & 3.3 & 8.0 & 3.6 & 8.0  \\
    Single-Modal + LM & 3.2  & 7.7  & 3.4 & 7.4  \\
    Multi-Modal  & 2.7 & 6.9 & 2.9 & 6.8 \\
    Multi-Modal  + LM & 2.6 & 6.6 & 2.9 & 6.6 \\
    \bottomrule
  \end{tabular}
\caption{Comparison between LM and multi-modal training}
\label{tab:lm-compare-result}
\end{table}

\vspace{-1em}

\subsubsection{Slot filling task}

To validate that the encoder trained under our proposed framework is richer in text information. We set up experiments on the SF task of SLU, and the results are shown in Table.\ref{tab:slu-result}. It shows the CERs of slot value and the F1s of slot type for the sequences generated by a transformer trained from scratch or the encoders of the transformer initialized by the parameters from a pretrained transformer on ASR task with single or multiple modalities. We use the encoder of a single modality transformer trained on \textsc{Librispeech} dataset for initialization, which is exactly the model for the approach named transformer from scratch in Table~\ref{tab:slu-result}. A significant improvement can be observed by pretraining the transformer encoder on a larger speech corpus. Initialization with multi-modality ASR model can yield a consistent improvement over that with single modality on SF task in both valid and test sets. With decoder initialization, the performance can be further improved. SF performance we achieved is comparable with those reported in ~\cite{yang2021superb}, which leveraged much larger size unlabeled data to get better-generalized representations for SLU.

\begin{table}[htbp]
  \centering
  \begin{tabular}{l  | c c | c c}
    \toprule
    \multirow{2}{*}{Initialization} & \multicolumn{2}{c|}{Valid} & \multicolumn{2}{c}{Test} \\
     & CER & F1 & CER & F1 \\
    \midrule
    Transformer from scratch & 55.80 & 67.30 & 57.91 & 67.14 \\
    Transformer Single Init Enc & 33.03 & 90.01 & 32.85 & 89.35 \\
    Transformer Multi Init Enc & 27.91 & 92.83 & 26.03 & 91.80 \\
    \qquad $+$  Multi Init Dec & 25.77 & 94.29 & 25.01 & 92.85  \\
    \bottomrule
  \end{tabular}%
\caption{Comparison between different initialization approaches on \textsc{Snips} slot filling task. }
\label{tab:slu-result}
\end{table}


\vspace{-1em}
\section{Conclusions}
\label{sec:conclusions}
In this paper, we propose two approaches to optimize the alignment of the speech and language latent spaces under the multi-modality E2E ASR framework.
We introduce a learnable embedding aligner, which is a shared linear projection between text encoder and speech encoder trained by MLM loss and CTC at phoneme level, respectively. The speech and text embeddings are expected to be pushed closer to the same latent space by the embedding aligner.
We further enhance the speech-text embedding alignment with modality switch training, which randomly swaps speech and text embeddings based on the forced alignment results. Phoneme aware and phoneme unaware strategies are compared for MST. Finally, we validated that the semantic information is injected into the speech embeddings with our proposed approach by conducting experiments on SF task of SLU. Experiments show that our proposed method achieves a relative 14\% to 19\% WER reduction on \textsc{Librispeech} ASR task and an absolute 2.5\% to 2.8\% F1 score improvement on \textsc{Snips} SF task.

\label{sec:refs}

\bibliographystyle{IEEEbib}
\bibliography{strings,refs}

\end{document}